# Semi-Competing Risks on A Trivariate Weibull Survival Model

Cheng K. Lee
Department of Targeting Modeling
Insight & Innovation
Marketing Division
Wachovia Corporation
Charlotte, NC 28244

Jenq-Daw Lee
Graduate Institute of Political Economy
National Cheng Kung University
Tainan, Taiwan  70101
ROC

## SUMMARY
A setting of a trivairate survival function using semi-competing risks concept is proposed. The Stanford Heart Transplant data is reanalyzed using a trivariate Weibull distribution model with the proposed survival function.

KEY WORDS: semi-competing risks; trivariate Weibull; terminal event

## 1. INTRODUCTION

Fine, Jiang and Chappell [1] introduced the term "semi-competing risk" in which one event censors the other but not *vice versa*. In their article, the bivariate Clayton survival function was used to demonstrate the concept. Even before them, Li [2] has worked on the same concept using the bivariate Weibull survival model by Lu and Bhattacharyya [3].  In his dissertation, Li described the censoring event as "termination event" and, as did Fine, Jiang and Chappell [1], the bivariate survival function was divided into two components with lower wedge and upper densities. Epstein and Muñoz [4], Shen and Thall [5], and Dignam, Weiand and Rathousz [6] worked on the likelihood function of a bivarirate survival model with four types of censoring events. Different from those authors, in this article, a trivariate survival function for semi-competing risks with two-fatal and one non-fatal events is first constructed and then followed by the likelihood function.

In Section 2, the Stanford Heart Transplant Data is reconstructed for the analysis of semi-competing risks. In Section 3, the trivariate Weibull survival function is proposed followed by the likelihood function. Section 4 shows the results of the analysis of the data. Finally, the article is concluded with some discussion in Section 5.

## 2. DATA STRUCTURE

The Stanford Heart Transplant Data has been analyzed by Aitkin, Laird, and Francis [7] using the Weibull, lognormal, and piecewise exponential models with the consideration of pre-transplant and post transplant survival. They also surveyed the literature of analyzing the same data. In addition, the same data was studied by Miller and Halpern [8] using four regression techniques. The analyses by Noura [9] and Loader [10] are to find the change point of the hazard rate.

As Boardman [11] pointed out, this classical data set has been intensively studied in the past, and "the data set really does not have too much going for it". Therefore, the results of this article are not to be compared with previous studies. Instead, the main purpose of this article is analyzing the same data from a new approach using a trivariate Weibull model with the concept of semi-competing risks.

The Stanford Heart Transplant Data analyzed in this study is from the article by Crowley and Hu [12] in which they denote $T_1$ the date of acceptance to the study, $T_2$ the date last seen, and $T_3$ the date of transplantation. $T_2$ is less than or equal to the last day of the data collection or the last day of the study which was April 1, 1974. An individual was to experience three events when the individual was accepted to the study. The 3 events are death before transplant ($E_1$), transplant ($E_2$), and death after transplant ($E_3$). Therefore, $E_1$ and $E_3$ are terminal events or fatal events, and $E_2$ is an intermediate event.

Let $X_1$ be the time to $E_1$, $X_2$ be the time to $E_2$, and $X_3$ be the time to $E_3$. $X_1$, $X_2$, and $X_3$ begin at $T_1$ and are in days. The three events $E_1$, $E_2$, and $E_3$ are competing with each other for the occurrence to each individual. However, these three events must occur in some certain orders. When $E_1$ occurs first, neither $E_2$ nor $E_3$ will occur because $E_1$ is a terminal event. When $E_2$ occurs, $E_3$ may occur later but $E_1$ will not occur because $E_1$ and $E_2$ are defined mutually exclusive. When $E_3$ occurs, $E_2$ must occur first because $E_3$ is the event defined to occur after $E_2$. With these orders, an individual must fall into one and only one of the following four cases. First case, an individual experiences $E_1$, and, therefore, no possibility for the occurrence of $E_2$ or $E_3$. The individual is said to be uncensored due to $E_1$. In this case, $X_1 = T_2 - T_1$, and $X_2$ and $X_3$ do not exist. Second case, an individual experiences $E_2$ first, and then $E_3$ with no occurrence of $E_1$. The individual is said to be uncensored due to $E_2$ and $E_3$. In this case, $X_2 = T_3 - T_1$, $X_3 = T_2 - T_1$, and $X_3$ does not exist. Third case, an individual experiences only $E_2$ before the end of the study. The individual is said to be uncensored due to $E_2$, and censored due to $E_3$. In this case, $X_2 = T_3 - T_1$, $X_3 = T_2 - T_1$, and $X_1$ does not exist. Fourth case, an individual does not experience $E_1$, $E_2$ or $E_3$ before the end of the study. The individual is said to be censored due to $E_1$, $E_2$ and $E_3$. In this case, $X_1 = T_2 - T_1$, $X_2 = T_2 - T_1$, and $X_3 = T_2 - T_1$. The original data of Crowley and Hu contains 103 observations. After deleting 3 observations with 0 in $X_1$, $X_2$ or $X_3$, and 4 observations of transplant with no mismatch score, there are total 96 observations included in this study.

## 3. THE MODEL AND THE LIKELIHOOD FUNCTION

The Weibull distribution model is chosen for the marginal distribution of $X_1$, $X_2$, and $X3$ as the model was adopted by Aitkin, Laird, and Francis [7] and Noura [9]. In order to study the relation among the three random variables, the following trivariate Weibull survival function is derived using Clayton copula [13].

$$S_{X_1,X_2,X_3} = \left(S_{X_1}^{1-\theta} + S_{X_2}^{1-\theta} + S_{X_3}^{1-\theta} - 2\right)^{\frac{1}{1-\theta}} \qquad (1)$$

where $S_{X_1} = Exp\left(-\left(\frac{x_1}{\lambda_1}\right)^{\gamma_1}\right)$, $S_{X_2} = Exp\left(-\left(\frac{x_2}{\lambda_2}\right)^{\gamma_2}\right)$, $S_{X_3} = Exp\left(-\left(\frac{x_3}{\lambda_3}\right)^{\gamma_3}\right)$, $0 < \lambda_1, \lambda_2, \lambda_3 < \infty$, $0 < \gamma_1, \gamma_2, \gamma_3 < \infty$, and $1 \leq \theta < \infty$. $X_1, X_2,$ and $X_3$ are independent when $\theta = 1$. One of the features of the Clayton copula is that it allows positive and negative association between the random variables. To account for the effects of covariates, let $\lambda_1$ be the exponential function of age at acceptance and previous surgery, and let both $\lambda_2$ and $\lambda_3$ be the exponential function of age at acceptance, previous surgery and mismatch score. Note that only individuals receiving transplant had mismatch score. The parameters in the proposed trivariate survival model are to be estimated by maximizing the likelihood function. When each of the three events can censor and be censored by other events, the proposed trivariate Weibull survival model is one of the components in the likelihood function. That is the component accounts for individuals of lost-to-follow-up or being censored at the end of the study. However, when semi-competing risks exist with the orders discussed in Section 2, the survival function for individuals censored at the end of the study or lost to follow-up becomes

$$S_{X_1}(t) + S_{X_2,X_3}(t,t) + \int_t^\infty \left[\frac{\partial}{\partial x_3} S_{X_2,X_3}(x_2,x_3)\right]_{x_2=x_3} dx_3. \qquad (2)$$

The detailed derivation is in the Appendix. As did Lawless [14], the component in the likelihood for case 1 is the negative derivative of equation (2) with respect to $x_1$. The component for case 2 is the derivative of equation (2) with respect to $x_2$ and $x_3$. The component for case 3 is the negative derivative of equation (2) with respect to $x_2$. And, the component for case 4 is equation (2) itself.

Therefore, the likelihood function for the proposed trivariate survival function with the four cases is

$$L(\boldsymbol{\theta}) = \prod_{i=1}^N f_{X_1}\left(t_{x_{1i}}\right)^{p_i}$$

$$\times f_{X_2,X_3}\left(t_{x_{2i}}, t_{x_{3i}}\right)^{q_i}$$

$$\times \left(\left[-\frac{\partial}{\partial x_2} S_{X_2,X_3}(x_2,x_3)\right]_{x_2=t_{x_{2i}}, x_3=t_{x_{3i}}}\right)^{r_i}$$

$$\times \left(S_{X_1}(t_i) + S_{X_2,X_3}(t_i,t_i) + \int_{t_i}^\infty \left(\frac{\partial}{\partial x_3} S_{X_2,X_3}(x_2,x_3)\right)_{x_2=x_3} dx_3\right)^{1-p_i-q_i-r_i} \qquad (3)$$

where are $p$, $q$, and $r$ are event indices and $t$ denotes the survival time.

## 4. RESULTS

The estimates and their corresponding 95% confidence intervals of the parameters are in the following table. The asymptotic covariance matrix is approximated by the inverse of the negative Hessian.

| Parameter | Estimate | 95% Confidence Interval |
|---|---|---|
| $\theta$ | 1.677 | ( 1.147,  2.208) |
| age at acceptance ($X_1$) | 0.087 | ( 0.060,  0.114) |
| previous surgery ($X_1$) | -1.316 | (-5.527, 2.894) |
| $\gamma_1$ | 0.342 | ( 0.258, 0.425) |
| age at acceptance ($X_2$) | 0.076 | ( 0.061, 0.091) |
| previous surgery ($X_2$) | 0.196 | (-0.653, 1.045) |
| mismatch score ($X_2$) | -0.036 | (-0.532, 0.460) |
| $\gamma_2$ | 0.733 | ( 0.614, 0.852) |
| age at acceptance ($X_3$) | 0.131 | ( 0.098, 0.165) |
| previous surgery ($X_3$) | 1.993 | (-0.242, 4.228) |
| mismatch score ($X_3$) | 0.340 | (-0.787, 1.467) |
| $\gamma_3$ | 0.422 | ( 0.322, 0.523) |

The results indicate only the age at acceptance is significantly different from zero at significance level of 0.05 for $X_1$, $X_2$, and $X_3$. The overall association parameter $\theta$ is 1.677 that indicates $X_1$, $X_2$, and $X_3$ are not much correlated.

## 5. DISCUSSION

In this article, the Clayton trivariate Weibull survival model with Weibull marginals is applied to the Stanford Heart Transplant data. Due to the order of the occurrences of the three events, a new formation of the likelihood function is proposed. The correlation coefficients between pairs of random variables can be obtained explicitly or numerically. The work of this article can also be expanded to higher dimensions.

## ACKNOWLEDGE

The author thanks Mr. Daniel Warren Whitman for his proofreading this article.

## APPENDIX

The fourth factor in the likelihood function is the probability that an individual is censored at $T_2$, the date last seen. After the censoring, although is unobservable, the individual may experience $E_1$ only, or $E_2$ followed by $E_3$. Suppose the individual is censored at time $t$, then the probability for the occurrences of the three events is $\Pr(t < X_1) + \Pr(t < X_2 < X_3)$. $\Pr(t < X_1)$ is simply equal to $S_{X_1}(t)$.
And, $\Pr(t < X_2 < X_3) =$
$$\int_t^\infty \int_t^{x_3} f_{X_2, X_3}(x_2, x_3) \, dx_2 \, dx_3$$

$$= \int_t^\infty \left[ \int_t^\infty f_{X_2,X_3}(x_2,x_3) dx_2 - \int_{x_3}^\infty f_{X_2,X_3}(x_2,x_3) dx_2 \right] dx_3$$

$$= S_{X_2,X_3}(t,t) - \int_t^\infty \int_{x_3}^\infty f_{X_2,X_3}(x_2,x_3) dx_2 dx_3$$

Considering $\int_{x_3}^\infty f_{X_2,X_3}(x_2,x_3) dx_2$,

$$\int_{x_3}^\infty f_{X_2,X_3}(x_2,x_3) dx_2$$

$$= \int_0^\infty f_{X_2,X_3}(x_2,x_3) dx_2 - \int_0^{x_3} f_{X_2,X_3}(x_2,x_3) dx_2$$

$$= f_{X_3}(x_3) - \int_0^{x_3} \left( \frac{\partial}{\partial x_2} \left( \frac{\partial}{\partial x_3} F_{X_2,X_3}(x_2,x_3) \right) \right) dx_2$$

$$= \frac{\partial}{\partial x_3} F_{X_3}(x_3) - \left[ \frac{\partial}{\partial x_3} F_{X_2,X_3}(x_2,x_3) \right]_{x_2=x_3} \quad (x_2 = x_3 \text{ denotes that } x_2 \text{ is replaced by } x_3)$$

$$= \left[ \frac{\partial}{\partial x_3} \left( F_{X_3}(x_3) - F_{X_2,X_3}(x_2,x_3) \right) \right]_{x_2=x_3}$$

$$= \left[ \frac{\partial}{\partial x_3} \left( -S_{X_2,X_3}(x_2,x_3) \right) \right]_{x_2=x_3}$$

Note that $S_{X_2,X_3}(x_2,x_3) = 1 - F_{X_2}(x_2) - F_{X_3}(x_3) + F_{X_2,X_3}(x_2,x_3)$ ([2]) where $F_{X_2}$, $F_{X_3}$ and $F_{X_2,X_3}$ are, respectively, the cumulative density function of $X_2$, $X_3$, and $X_2$ and $X_3$.

Then, $\Pr(t < X_2 < X_3) = S_{X_2,X_3}(t,t) + \int_t^\infty \left[ \frac{\partial}{\partial x_3} S_{X_2,X_3}(x_2,x_3) \right]_{x_2=x_3} dx_3$.

Therefore, the fourth factor in the likelihood function is

$\Pr(t < X_1) + \Pr(t < X_2 < X_3) = S_{X_1}(t) + S_{X_2,X_3}(t,t) + \int_t^\infty \left[ \frac{\partial}{\partial x_3} S_{X_2,X_3}(x_2,x_3) \right]_{x_2=x_3} dx_3$.